\documentclass[prb,twocolumn,showpacs]{revtex4}
\usepackage{graphicx}
\begin{document}

\title{Theoretical Study of Spin-dependent Electron
Transport in Atomic Fe Nanocontacts}

\author{Hugh Dalgleish and George Kirczenow}

\affiliation{Department of Physics, Simon Fraser
University, Burnaby, British Columbia, Canada V5A 1S6}

\date{\today}

\begin{abstract}

We present theoretical predictions of spintronic
transport phenomena  that should be observable in
ferromagnetic Fe nanocontacts bridged by chains of Fe
atoms.  We develop appropriate model Hamiltonians based
on  semi-empirical considerations and the known
electronic structure of bulk Fe derived from {\em ab
initio} density functional calculations. Our model is
shown to provide a satisfactory description of the
surface properties of Fe nano-clusters as well as bulk
properties.   Lippmann-Schwinger and Green's function
techniques are used together with Landauer theory to
predict the current, magneto-resistance, and spin
polarization of the current in Fe nanocontacts bridged by
atomic chains under applied bias.  Unusual device
characteristics are predicted including negative
magneto-resistance and spin polarization of the current,
as well as spin polarization of the current for
anti-parallel magnetization of the Fe nanocontacts under
moderate applied bias.  We explore the effects that
stretching the atomic chain has on the magneto-resistance
and spin polarization and predict a cross-over regime in
which the spin polarization of the current for parallel
magnetization of the contacts switches from negative to
positive.  We find resonant transmission due to dangling
bond formation on tip atoms as the chain is stretched
through its breaking point to play an important role in
spin-dependent transport in this regime. The physical
mechanisms underlying the predicted phenomena are
discussed.

\end{abstract} 


\maketitle

\section{Introduction}

Spin-dependent transport (SDT) phenomena occur when a
bias voltage is applied across a junction between
materials one or both of which are
magnetic.\cite{reviews} They include partial or complete
spin-polarization of the electric current and changes in
the electrical resistance when the magnetization
direction of one of the magnetic components of the system
is reversed through the application of a magnetic field.
The latter effect is usually referred to as ``junction
magneto-resistance" or ``giant magneto-resistance" (GMR).
SDT has been observed at interfaces  between
ferromagnetic metals and superconductors
\cite{Meservey1970}, between ferromagnetic and normal metals
\cite {Johnson1985}, between ferromagnetic metals separated
by thin insulating films \cite{Julliere1975}, and more
recently between magnetic semiconductors and nonmagnetic
semiconductors.\cite{Fiederling1999,Ohno1999}
 
At the present time spin-dependent transport through
nanoscale junctions is attracting increasing attention. 
SDT through molecules bridging nanoscale  magnetic
contacts has been investigated
theoretically\cite{Emberly2002,Pati2003,
Babiaczyk2004,Rocha2005}
and relevant experiments have recently been
reported.\cite{Dediu2002,Xiong2004,Petta2004,Wu2005} 
SDT through chains of atoms
or single atoms connecting pairs of nickel and cobalt
nanocontacts has also been investigated
theoretically.\cite{Tatara1999,Velev2004,
Bagrets2004,Palacios2004,Rocha2004}
Such systems have been realized experimentally using
break-junction\cite{Oshima1998,Garcia1999,
Ono1999,Viret2002,
Untiedt2004} 
and electrochemical\cite{Sullivan2004} techniques
and transport measurements on them have been carried
out.\cite{Garcia1999,Viret2002,
Oshima1998,Ono1999,
Untiedt2004,Sullivan2004}
SDT measurements have also been carried out on Fe atomic
contacts and magnetoresistances an order of magnitude
smaller than for the Ni and Co systems were
found.\cite{Garica2000} This difference was
explained in terms of the smaller ratio between the spin
$up$ and spin $down$ densities of states at the Fermi
level in bulk Fe than in Ni or Co.\cite{Garica2000}
Very recently, a theoretical study elucidating the
systematics of SDT through a variety of nanocontacts,
including Fe nanocontacts connected by atomic chains, has
been reported.\cite{Velev2004}  The
calculations were based on a semi-empirical tight-binding
model and were carried out in the limiting case of
infinitesimal applied bias voltage. The structures
studied were periodic arrays of nanoconstrictions with
all of the Fe atoms occupying sites of a bulk crystal
lattice and enough vacant sites being included in the
structure to isolate the Fe chains from each other. 

In this article we explore SDT through Fe atomic chains
connecting Fe nanocontacts theoretically in some
important regimes that were not considered in the
previous work: We develop a model applicable to more
general atomic geometries; thus we are able to examine
the effects on SDT of stretching the atomic chain that
bridges the Fe nanocontacts as occurs in break-junction
experiments. We predict pronounced resonant
spin-dependent transport phenomena due to dangling bonds
that form on the tip atoms as the atomic chain parts. We
also consider the application of finite bias across the
junction of the Fe nanocontacts and predict other unusual
SDT phenomena including negative junction
magneto-resistance and  negative spin-polarization of the
current. We define these quantities as follows: 

The junction magneto-resistance (JMR) is defined as 
\begin{equation}
JMR=\frac{(I_{par}-I_{anti})}{\frac{1}{2}(I_{par}+I_{anti})}
\label{JMR}
\end{equation}  where $I_{par}$ ($I_{anti}$) is the
electric current flowing between the Fe nanocontacts when
their magnetizations are parallel (antiparallel). 

The spin-polarization of the current (SP) is defined as
\begin{equation}
SP=\frac{(I_{up}-I_{down})}{\frac{1}{2}(I_{up}+I_{down})}.
\label{SP}
\end{equation} Note that we refer to the majority spins
in {\em each} contact as ``spin $up$" and the minority
spins as ``spin
$down$".  Thus if an electron is transmitted between
contacts with anti-parallel magnetizations without
changing its spin orientation, the transition will be
referred to as a spin $up$ $\rightarrow$ spin $down$ or
spin $down$
$\rightarrow$ spin
$up$ transition.  In equation (\ref{SP}) $I_{up}$
($I_{down}$) is the spin $up$ (spin $down$) current with
$up$ and
$down$ defined as for the nanocontact that is the
electron {\em drain} electrode for the system. The total
current is defined as
$I=I_{up}+I_{down}$ for both parallel and antiparallel
magnetizations. 

We find that, interestingly, in equation (\ref{SP}) the
current for the spin with the larger bulk density of
states is not always larger than the current for the spin
with the smaller density of states. Thus the sign of the
spin polarization of the current cannot be predicted from
a knowledge of the spin $up$ and $down$ densities of
states alone, even when the magnetizations of the two
nanocontacts are parallel.  We also show that the spin
polarization of the current need not vanish even for the
case of anti-parallel magnetizations of the two
nanocontacts because of a symmetry breaking that occurs
in the system under finite applied bias. Furthermore we
predict that the sign of the the junction
magnetoresistance should change when the atomic chain
connecting the nanocontacts is stretched and also if the
applied bias voltage is increased sufficiently. 

Our SDT calculations are based on Landauer
theory\cite{Dattasbook} and Lippmann-Schwinger and
Green's function techniques.  Transport calculations
require a knowledge of the underlying electronic
structure. Semi-empirical tight-binding models of
electronic structure have been used successfully in
modeling SDT in thin film structures involving Fe and
other magnetic metals together with insulating or vacuum
tunnel barriers or non-magnetic metal
spacers,\cite{Tsymbal1997,Mathon2001,Mathon2002,
Mathon1997,MathonUmerski1997}
and also SDT in magnetic metal
nanocontacts.\cite{Velev2004} They have also
been used successfully to explain the experimental
current-voltage characteristics of molecular nanowires
connecting non-magnetic metal
electrodes.\cite{Datta1997,Emberly2001,Kushmerick2002} In this
paper, we model the Fe nanocontacts as nanoclusters of Fe
atoms connected to ideal leads that represent the source
and drain electrodes.   However we parameterize the
electronic structure of the Fe clusters with the use of
semi-empirical tight-binding parameters obtained from
fitting the known band structure of magnetic bulk
Fe.\cite{Pap}   Never-the-less, as we explain in Section
\ref{Model}, our model also provides a satisfactory
description of the spin-resolved surface densities of
states and of the local magnetic moments at the surfaces
of Fe nanoclusters. Thus our model incorporates in an
approximate way $both$ the bulk and surface magnetic
properties of Fe which together influence the SDT through
a junction of bulk Fe leads that come together at a
nanocontact, as in experimental realizations of Fe atomic
chains.  

This article is organized as follows: In Section
\ref{Model} we describe our model of the geometry and
electronic and magnetic structure of the Fe nanocontacts.
In Section \ref{Theory} we summarize the formalism used
in our spin-dependent transport calculations.  In Section
\ref{Results} this formalism is applied to different
arrangements of Fe atoms connecting the ferromagnetic
nanoclusters: We start by considering a structure in
which the positions of the atoms of the Fe nanocontacts
and of the atomic bridge connecting them coincide with
sites of an Fe crystal, however we consider a single
atomic bridge, as distinct from the periodic array of
nanocontacts treated in Ref.
\onlinecite{Velev2004}.  The spin-dependent
currents flowing through this junction are calculated
utilizing initially a simple, voltage-independent model
of the transmission probabilities of spin
$up$ and spin $down$ electrons through the junction. A
strong magneto-resistance is predicted.  We then perform
voltage-dependent calculations of the transmission to
examine the effects of finite bias on the spin-dependent
predictions more closely and obtain a still strong,
though relatively weaker, JMR.  We then proceed to
examine geometries in which the separation of
nanoclusters is allowed to vary, stretching the atomic
chain.  To this end, we supplement the above
tight-binding model of the electronic structure of the Fe
clusters\cite{Pap} by introducing a position-dependent
parameterization for the electronic coupling between
clusters, in the spirit of extended-H{\"u}ckel theory. 
We allow the separation between the tips of the clusters
to range from a  bulk nearest-neighbor distance, in which
case ballistic SDT is predicted, to a distance at which
the atomic chain has been broken and
vacuum-tunneling-like SDT occurs.   The cross-over regime
is examined and interesting phenomena are predicted,
including spin-dependent transport resonances mediated by
dangling bonds on the tip atoms.     In the ballistic
regime, unusual device characteristics such as negative
JMR and spin polarization of the current are predicted.
We then proceed to apply our approach to the case of
tunneling between ferromagnetic nanoscale tips separated
by vacuum and examine the spin-dependent current and
magneto-resistance. Physical mechanisms are presented
that explain the predicted effects. The
$I-V$ characteristics and magneto-resistance predictions
presented in this paper should be experimentally
accessible. Our conclusions are summarized in Section
\ref{Conclusion}.

\section{The Model}
\label{Model}

The system of interest consists of two bulk ferromagnetic
metal electrodes that act as a source and drain for
electrons, joined by a nanoscale junction of atoms of the
same magnetic metal. We model  this structure as a
connected pair of nanoscale contacts of (100)
body-centered cubic iron, each contact consisting of a 55
atom cluster built from 5x5, 4x4, 3x3, 2x2 layers of
atoms, terminated with a single tip atom.  Thus the
atomic chain bridging the nanocontacts is a dimer
consisting of the two tip Fe atoms.

The electronic structure of the Fe clusters is described
by a tight-binding Hamiltonian  using a non-orthogonal
basis of $s$, $p$ and
$d$ atomic orbitals, a total of 9 orbitals per atom for
each spin orientation. The values of the Hamiltonian
matrix elements
$H_{i,j}$  and overlaps $S_{i,j} = \langle i | j \rangle$
between orbitals $i$ and $j$ are,  where possible,  taken
from Ref.\onlinecite{Pap}. These tight-binding parameters
are based on fits to {\em ab initio} band structures
calculated for Fe crystals
\cite{Pap} and have previously been employed successfully
to study magnetic multilayer systems \cite{Mathon2001} and
magnetic\cite{Velev2004} and
non-magnetic\cite{Cuevas1997} atomic chain nanoscale
systems.  However this parameter set is modified and/or
extended appropriately (as discussed below and in Sections
\ref{nearest neighbor} and \ref{var geom}, respectively)
when a magnetic domain wall is present or a finite bias
voltage is applied across the junction or the positions
of the atoms of the junction do not all  coincide with
sites of a perfect bcc Fe crystal lattice. 

The source and drain electrodes are modeled as sets of
one-dimensional ideal leads coupled to each cluster.
These ideal leads are implemented as semi-infinite 
tight-binding chains with one orbital per site, one such
lead being coupled to each ($s$,$p$ and $d$, spin $up$
and spin $down$) orbital of each atom of the two layers
of each Fe cluster that are furthest from the junction.
The site-orbital energy of each ideal lead is matched to
that of the Fe tight binding orbital to which the lead
couples and the lead band widths are chosen large enough
that all of the eigenstates of the leads are propagating
modes in the energy range of interest. As well as
mimicking macroscopic electrodes by supplying an ample
electron flux to the system, this large number of ideal
source and drain leads (369 per cluster for each spin
orientation) has an effect similar to phase-randomizing
B{\"u}ttiker probes\cite{Buttiker1986} in minimizing
the influence of dimensional resonances due to the finite
sizes of the Fe clusters employed in the model. Thus our
results do not change qualitatively if the parameters of
the ideal leads are varied (within reasonable bounds) or
an additional layer of Fe atoms is included in the
clusters. 

We assume that each cluster (including the tip atom) and
the semi-infinite leads connected to it form a single
magnetic domain. Thus, if the magnetizations of the two
nanocontacts are anti-parallel, an atomically thin, hard
domain wall is present at the midpoint of the Fe dimer
connecting the two clusters, in accordance with the
calculation of Ref.
\onlinecite{Bruno1999} and with previous work modeling
SDT in Ni atomic contacts.\cite{Palacios2004} The
tight-binding parameters developed in Ref.
\onlinecite{Pap} do not include the off-diagonal
Hamiltonian or overlap matrix elements between atomic
orbitals located on opposite sides of a magnetic domain
wall. However the published off-diagonal matrix elements
connecting spin $up$ orbitals are not very different from
the corresponding matrix elements connecting spin $down$
orbitals in the same domain.\cite{Pap} This is physically
reasonable given the that spatial part of a spin $up$
orbital should be similar to that of the corresponding
spin $down$ orbital. Thus in the present work, at zero
bias, when geometrically applicable, we approximate the
off-diagonal matrix elements connecting orbitals that are
on {\em opposite} sides of an atomically thin domain wall
and have the {\em same} spin orientation by the average
of the spin $up$ and spin
$down$ matrix elements connecting the corresponding
orbitals in a single domain. As in in Ref.
\onlinecite{Pap},  we also take all Hamiltonian and
overlap matrix elements connecting orbitals having
opposite spin orientations to be zero,  i.e., spin flips
during electron  transmission through the junction are
not considered. 

\begin{figure}
\includegraphics[width=0.99\columnwidth]{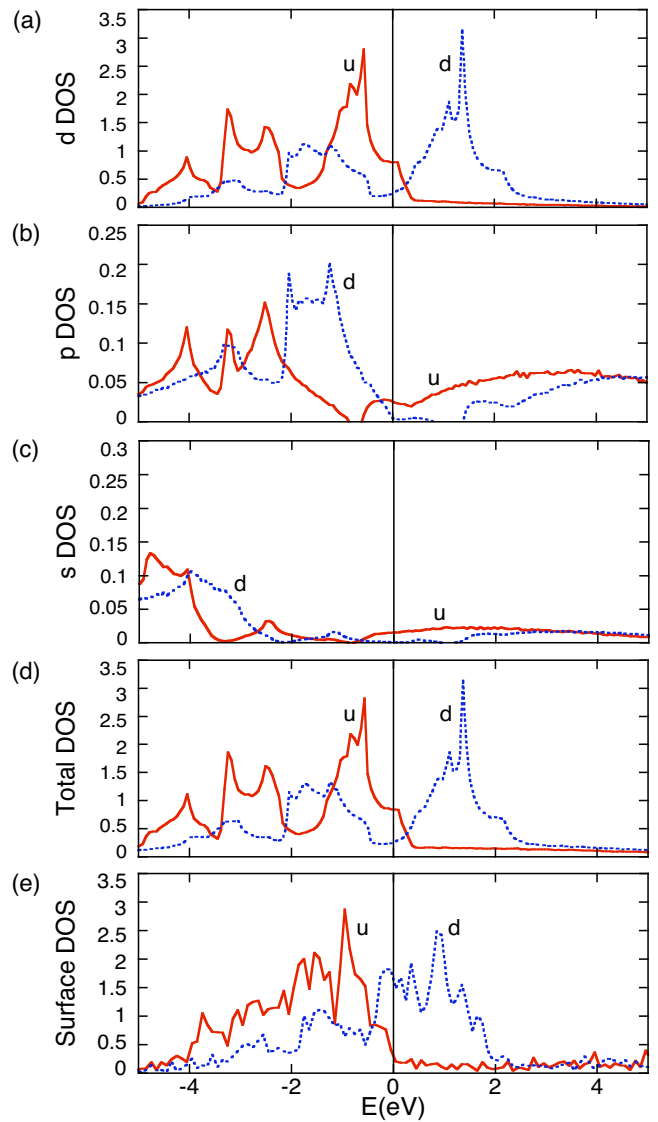}
\caption{(Color online) Calculated bulk (a-d) and surface (e)
densities of states vs. energy (eV) at zero bias for spin
$up$ (solid curve u) and spin
$down$ (dotted curve d).  The Fermi energy is located at
0eV. (a) The densities of states due to
$d$-electrons. (b) The densities of states due to
$p$-electrons. (d) The densities of states due to
$s$-electrons. (d) The total densities of states. 
The $d$ orbital contribution dominates the
total density of states except at the higher and
lowest energies shown.
(e) The
surface contribution to the densities of states.}
\label{Fig_1}
\end{figure}
  
To better understand the significance of the results of
our SDT calculations it will be useful to compare them to
particular features of the spin-resolved surface and bulk
densities of states of the Fe electrodes and to further
resolve the density of states into its
$s$,
$p$ and
$d$ components. Since the atomic orbital basis that we
use is non-orthogonal we  carry out this resolution as
follows: Consider the normalized eigenstate
$|\Psi_{k}\rangle = \sum_{i} c_{k,i} |i\rangle$ of the
Hamiltonian, where $|i\rangle$ is an atomic orbital.
Express the norm of the eigenstate in terms of the
contributions of all of the atomic orbitals of the system
as
\begin{equation}
\langle \Psi_{k}|\Psi_{k}\rangle = \sum_{i,j} c^{*}_{k,i}
c_{k,j} S_{i,j}
\label{Mull2}
\end{equation} where $S_{i,j} = \langle i | j \rangle$.
To resolve the contribution of eigenstate
$|\Psi_{k}\rangle$ to the density of states into its $s$,
$p$ and $d$ orbital components we assign to each atomic
orbital $i$ the Mulliken weight
$c^{*}_{k,i} c_{k,i} + \sum_{j \ne i} (c^{*}_{k,i} c_{k,j}
S_{i,j}+c^{*}_{k,j} c_{k,i} S_{j,i})/2$. The partial and
total bulk densities of states obtained in this way are
shown in Fig. \ref{Fig_1} (a-d).

In order to assess the applicability of our
semi-empirical tight binding model (that is based on fits
to bulk band structures\cite{Pap}) to nanostructures that
include surfaces, we calculated the local magnetic moment
per atom in the surface atomic layer and in the interior
of each of several large Fe clusters.  A similar Mulliken
analysis to that described above was used to resolve the
electron probability distributions for individual
eigenstates of the tight-binding Hamiltonian used in the
present work into surface and interior contributions. We
found the lower coordination number and lack of symmetry
at the surface to result in enhanced magnetic moments at
the surface, approximately 2.5$\mu_{B}$ per atom at the
surface versus 2.2$\mu_{B}$ in the bulk for our model, a
result qualitatively similar to that of {\em ab initio}
surface calculations for Fe (2.25 in the bulk and 2.98 at
the surface).\cite{Ohnishi1983} We note that Mulliken
analyses arbitrarily assign half of the probability
contribution that is due to overlaps between two atoms in
equation (\ref{Mull2}) to each of the two atoms involved
and therefore tend to underestimate differences between
the electronic populations of adjacent atoms, often
significantly. Thus we regard the level of agreement
between the present model and the {\em ab initio} surface
calculations\cite{Ohnishi1983} as satisfactory.  The
contributions from the surface layer of a representative 
Fe cluster to spin {\em up} and spin {\em down} densities
of states obtained in this way are shown in Figure
\ref{Fig_1}(e) and compare well with those reported in
Ref. \onlinecite{Ohnishi1983}.

\begin{figure}
\includegraphics[width=0.99\columnwidth]{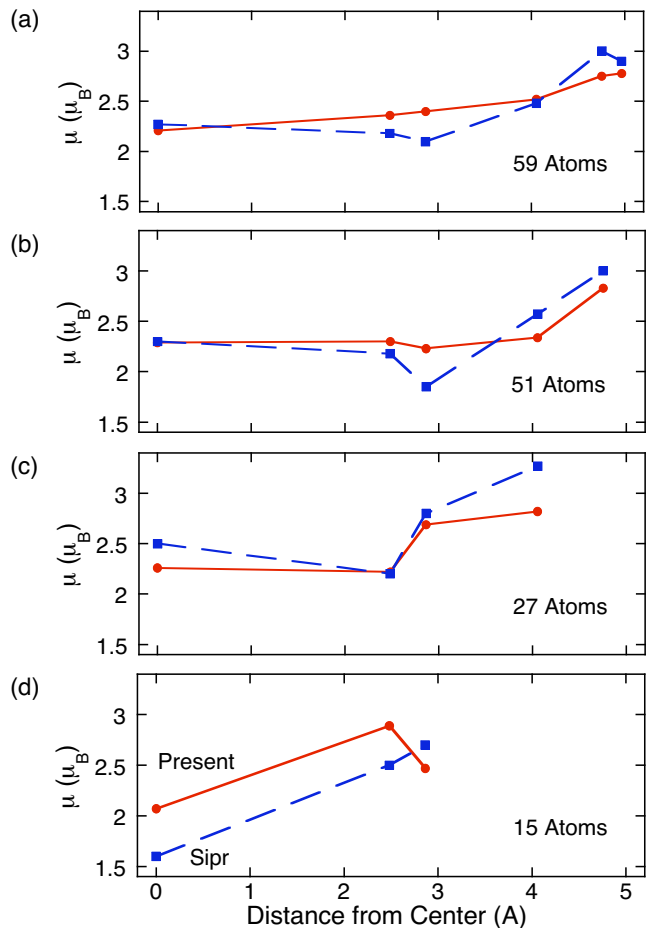}
\caption{(Color online) Calculated magnetic moments per atom as a 
function of atom distance relative to the central atom of 
Fe nanoclusters.  Solid lines correspond to the present 
study, dashed lines to the first principles calculations of \v{S}ipr 
{\em et al}.\cite{Sipr2004}  All atoms are at bulk 
lattice positions and cluster size is increased as
additional coordination shells are added.  (a) 15 atom 
Fe cluster.(b) 27 atom cluster. (c) 51 atom cluster. 
(d) 59 atom cluster.}
\label{Fig_2}
\end{figure}

As further indication that our semi-empirical model is 
appropriate for describing also the nanocontact region 
(the two tip atoms) in addition to the rest of the 
contacts approximating bulk electrodes, we have employed 
the model to predict magnetic moments of {\em actual} Fe 
nanoclusters of increasing size.  The results of this 
calculation are compared to those of {\em ab initio} 
calculations on identical nanoclusters 
\cite{Sipr2004} in Figure \ref{Fig_2}(a-d).  It is  
evident that the predicted moments of our semi-empirical 
model compare well to those of the {\em ab initio} study 
for both the surface and interior atoms of the clusters, 
particularly when the cluster size resembles that of our 
nanoclusters used for the transport calculation (51 and 
59 atom clusters compared to the 55 atom clusters we 
employ in our following study).  We consider the good 
agreement for the outermost atoms of the clusters as 
indication that our model adequately reproduces the 
magnetic properties of even atoms of low 
coordination number, such as the tip atoms of the clusters in our 
transport calculations.  Additionally, as the nanocluster 
size is increased (to the 51 and 59 atom sized 
clusters), the magnetic moments of all of the interior atoms 
quickly approach that of the bulk.  Therefore these 
results also provide further evidence that the Fe clusters
employed in our transport calculations are large enough
to model macroscopic magnetic Fe electrodes (with single atom 
tips) if our
semi-empirical tight binding model is used to treat the
electronic and magnetic structures of these systems. 

\section{Theory of Spin-Dependent Transport}
\label{Theory}       

When a bias $V$ is applied between the ferromagnetic Fe
electrodes an electric current $I$ flows through the
nanoscale junction described in Section \ref{Model}. 
Landauer theory\cite{Dattasbook} relates this current to
the multi-channel probability
$T$ for an electron to scatter from the source electrode
to the drain via the junction, according to 
\begin{equation} I(V) = \frac{e} {h} \int dE\:
T(E,V)[f(E,\mu_{S})-f(E,\mu_{D})]
\label{LandauerEQ}
\end{equation} where $E$ is the energy of the electron,
$F(E,\mu)$ is the equilibrium Fermi distribution and
$\mu_{S,D} = E_{F}
\pm eV/2$ are the electro-chemical potentials of the
source (S) and drain (D) electrodes in terms of the
common Fermi energy,
$E_{F}$.\cite{zeroT} The transmission probability in
equation (\ref{LandauerEQ}) is 
\begin{equation} T(E,V) = \sum_{\alpha,\beta,s,s'}
|\frac{\nu_{\beta,s'}} {\nu_{\alpha,s}}|
|t_{\beta,s';\alpha,s}|^{2}
\label{trans}
\end{equation} where $t_{\beta,s';\alpha,s}$ is the
transmission amplitude from a state of ideal lead
$\alpha$ of the source electrode with spin $s$  to a
state of ideal lead $\beta$ of the drain electrode with
spin $s'$ and
$\nu_{\alpha,s}$ and $\nu_{\beta,s'}$ are the
corresponding electron velocities. If spin flips during
transmission of electrons through the junction are
neglected as in the present work and the summations over
spin in equation (\ref{trans}) for $T$ are restricted to
a particular spin orientation then  equation
(\ref{LandauerEQ}) yields the current for that spin
orientation, i.e., $I_{up}$ or $I_{down}$.  

We calculate $T$ numerically by solving the
Lippmann-Schwinger equation\cite{nonorthogbasis}
\begin{equation} |\Psi^{\alpha}\rangle =
|\Phi^{\alpha}_{0}\rangle + G_{0}(E) W
|\Psi^{\alpha}\rangle
\label{LSeq}
\end{equation}  where $G_{0}(E)$ is the Green's function
for the system with the coupling $W$ between the ideal
leads and Fe clusters switched off,
$|\Phi^{\alpha}_{0}\rangle$ is the eigenstate of ideal
lead 
$\alpha$ when it is decoupled from the Fe clusters, and
$|\Psi^{\alpha}\rangle$ is the corresponding  eigenstate
of the complete system with coupling $W$ switched on. 
The transmission amplitudes
$t_{\beta,s';\alpha,s}$ that appear in equation
(\ref{trans}) are obtained from the solution
$|\Psi^{\alpha}\rangle$ of the Lippmann-Schwinger
equation  (\ref{LSeq}).

\section{Results}
\label{Results}

\subsection{Fe Nanocontacts Bridged by Fe Atoms in a bcc
Nearest-Neighbor Geometry}
\label{nearest neighbor}

\begin{figure}
\includegraphics[width=0.99\columnwidth]{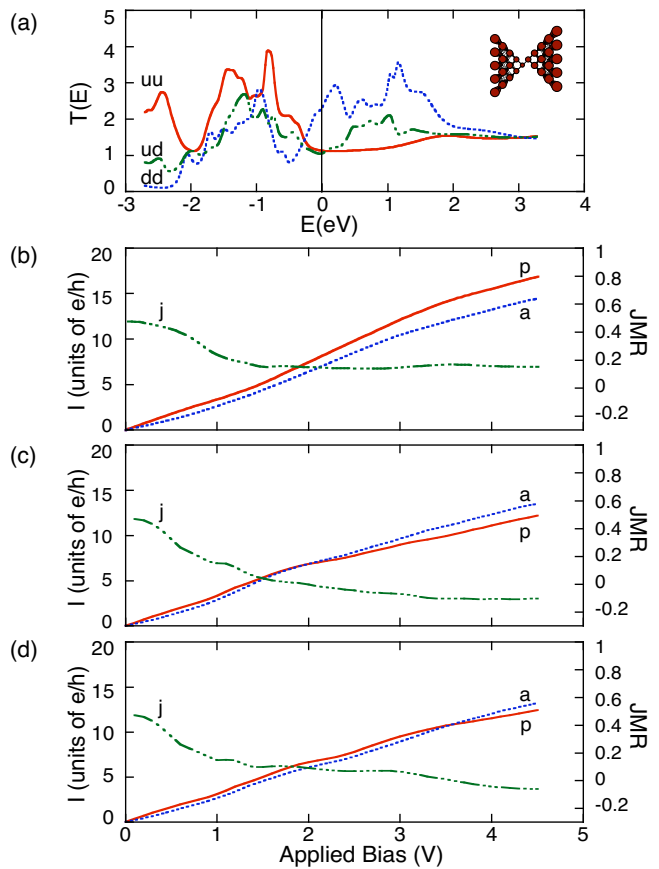}
\caption{(Color online) (a) Transmission probabilities as a function of
energy (eV) at zero bias for the contact geometry  shown
in the inset where all of the atoms are located at sites
of a bulk Fe bcc lattice. Spin
$up \rightarrow up$ (uu), $down \rightarrow down$ (dd)
transmission probabilities are for parallel magnetization
of the two contacts;
$up \rightarrow down$ (ud) is for anti-parallel
magnetization. The Fermi energy is located at 0eV. (b)
The current as a function of voltage for parallel (p) and
anti-parallel (a) magnetizations calculated by
integrating the zero bias transmission probabilities in
(a).  The JMR (j) is also shown. (c) The currents and JMR
as calculated with an applied bias using the linear
voltage drop model. (d) The currents and JMR as
calculated with an applied bias using the abrupt voltage
drop model.}
\label{Fig_3}
\end{figure}

We begin by considering the nanocontact geometry shown in
the inset of Figure \ref{Fig_3}(a). Here the two 55 atom
clusters are placed so that the atom terminating the tip
of each cluster is at  a bulk nearest neighbor position
relative to the tip atom of the other cluster. Thus all
atoms occupy positions that match sites of a
body-centered cubic Fe bulk lattice so that  a
tight-binding Hamiltonian and overlap matrix for the
entire structure can be constructed using the parameter
set of Ref.
\onlinecite{Pap}. Figure
\ref{Fig_3}(a) shows the calculated transmission
probabilities for the different spin configurations, at
zero bias, for this geometry. The spin
$up \rightarrow up$ (uu) and  $down \rightarrow down$
(dd) transmission probabilities are for parallel
magnetization of the two contacts;
$up \rightarrow down$ (ud) is for anti-parallel
magnetization. The Fermi energy is at 0eV. The
transmission characteristics depend somewhat on the
specific details of the chosen geometry, such as cluster
size and shape.  However, quite similar results are 
obtained when a 6x6 atom layer is added to each of the
clusters.  Current and magneto-resistance results are 
also robust to geometrical changes, such as addition of a
6x6 layer, or deletion of the 5x5 layer that terminates
the nanocluster, indicating that the clusters and ideal
leads in our model adequately represent real
(macroscopic) Fe leads.  

For this geometry, the calculated spin $up \rightarrow
up$ transmission, perhaps surprisingly for a transition
metal, exhibits a fairly flat plateau close to unity near
and above the Fermi energy.  As for atomic chains
constructed from some non-magnetic metals
\cite{Cuevas1997}, this transmission near unity in the
spin $up$ case is not due to a single open conducting
channel, but to the superposition of partially conducting
channels.  By systematically switching off the tip
atom-to-tip atom coupling for different atomic orbital
symmetries, we deduce that the $up
\rightarrow up$ transmission near the Fermi energy is due
primarily to
$s$ and $p$ electrons, even though as is seen in Figure
\ref{Fig_1} the spin $up$ density of states due to $d$ 
electrons is
comparable to those due to $s$ and $p$ electrons above
the Fermi energy.  Below the Fermi energy, the bulk and 
surface densities of states for $up$ electrons are 
larger (primarily $d$-like), and the predicted 
transmission is also higher. 

In contrast to the spin $up$ case, the spin $down
\rightarrow down$ transmission near the Fermi energy, not
only shows a strong contribution from $p$-type electrons
but a very strong $d$-electron component;
$s$-electrons do not seem important in this case.
This also can be contrasted to the spin $down$ densities 
of states very near the Fermi energy
where that due to $p$-electrons is small and comparable
to that for $s$-electrons.
Yet the presence of many common features that can be observed 
in the calculated
transmission and the surface and bulk densities of states 
does support the conclusion that the spin
transport properties of this magnetic system are
influenced by the densities of states.  However, the
transmission characteristics can not be determined from
solely a knowledge of the densities of states;
characteristics such as the near quantum conductance
above the Fermi energy in the spin $up$ channel and the
magnitudes of the transmission features can not be
deduced from the densities of states alone.  Those
characteristics can be determined only from a full
quantum transport calculation such as that presented
here. 

Interestingly, as shown in Figure \ref{Fig_3}(a), the
calculated spin
$down \rightarrow down$ transmission  is significantly
larger than the spin $up \rightarrow up$ transmission
near the Fermi energy and thus  according to equations
(\ref{SP}) and (\ref{LandauerEQ}) our theory predicts a
negative spin polarization  of the current through this
junction at low bias for parallel magnetization of the
two contacts.  Negative spin polarization of the current
is a common feature  of theories that attempt to estimate
spin-dependent transport effects in Ni and Co
heterostructures simply from convolutions of the
spin-dependent densities of states of the magnetic
electrodes that are dominated by the contributions of $d$
electrons.  However, such approximations have been shown
incorrect in magnetic tunnel junctions, as
$d$-electron wave functions decay very quickly in the
insulating barrier
\cite{Meservey1970,Mathon2001,Mathon2002,Mathon1997,
Butler2001,Tsymbal2003}. 
This decay is further enhanced by the conservation of
wavevector parallel to the interface
\cite{Butler2001,MacLaren1997}, and
therefore the
$d$-electrons that are responsible for the strongest
features in the densities of states do not play an active
role in transmission in such insulator mediated systems. 
In atomic chain systems however, there is no lateral
periodicity, and parallel wavevector conservation is
broken.  Also, the magnetic electrodes that we consider
are in this case in physical contact and so
$d$-electrons do play an active role in transport
\cite{dtransportGarcia}.  Never-the-less, a convolution
of the bulk densities of states cannot account for the
negative spin polarization of the current that we predict
in the present system;  it arises from a combination of
effects due to the surface and bulk densities of states
{\em and} our full quantum mechanical treatment of
electron transmission through the junction.

The transmission probability $T(E,V)$ that enters the
Landauer expression (\ref{LandauerEQ}) for the electric
current $I$ flowing through the nanocontact depends on
the applied bias voltage $V$ as well as the electron
energy. The bias dependence of $T$ depends on the
potential profile through the nanocontact which is
difficult to calculate from first principles since this
is a non-equilibrium many-body property. However,
appropriate heuristic models for the profile can yield
accurate results for the current.\cite{Ke2004} We
adopt this approach here by comparing the results
obtained for a variety of simple models.   

Figure \ref{Fig_3}(b) shows the current calculated by
approximating
$T(E,V)$ with $T(E,0)$, an approximation commonly used to
study transport at low values of the bias. Here, and in
all of the figures and discussion that follow, the 
current refers to the flow of electrons, or electron flux, 
as electrons are transported from the source 
to the drain electrode.  In Figure \ref{Fig_3}(b) the
calculated current is higher for parallel magnetization
than for anti-parallel magnetization and steadily
increases with applied bias, characteristic of ballistic
transmission.  The calculated JMR, as defined by Eq.
\ref{JMR} is positive and larger at low bias, decaying to
a lower value of about 0.15 at high bias.  

Figure \ref{Fig_3}(c) and (d) show our results obtained
using explicitly bias-dependent transmission
probabilities $T(E,V)$ that were calculated for two
different models of the potential profile of the
nanocontact. In each case we assume that the entire
voltage drop occurs across  the narrowest constriction in
the system, i.e, over the dimer.   Thus all atoms (and
ideal leads) to the source side of the dimer are assumed
to be at a potential
$\phi = -\frac{V}{2}$, while atoms and leads to the drain
side are at
$+\frac{V}{2}$.  For the results shown in Figure
\ref{Fig_3}(c) the potential is assumed to vary linearly
through the region occupied by the dimer. 
\cite{Ke2004,Pernas1990,Mulica2000,
Damle2001,Pleutin2002}
In Figure \ref{Fig_3}(d),  for comparison, the potential
is assumed to change abruptly  from $\phi = \pm
\frac{V}{2}$ in the atomic layers adjacent to the dimer
to $\phi = 0$ on the two atoms that constitute the dimer
itself, a profile analogous to that proposed initially in
theoretical work on ballistic semiconductor
nanostructures\cite{Glazman1989} and more recently adopted in
modeling certain  molecular
wires.\cite{EmberlyKirczenow1998,Mulica2000}

In each case the electrostatic potential modifies the
diagonal matrix elements of the  tight-binding
Hamiltonian of the system which become
$H_{i,i} = H^0_{i,i} -e \phi_i$ where $H^0_{i,j}$ is the
tight-binding Hamiltonian matrix at zero bias and 
$\phi_i$ is the electrostatic potential $\phi$ at the
site occupied by atomic orbital $i$. Because the tight
binding basis that we use is non-orthogonal, the applied
electrostatic potential also modifies the non-diagonal
Hamiltonian matrix elements.\cite{shift} Here we include
this effect approximately in the form 
\begin{equation} H_{i,j} = H^0_{i,j}-eS_{i,j}(\phi_i +
\phi_j)/2 .
\label{electroshift}
\end{equation}

The applied electrostatic potential breaks the symmetry
between the left and right clusters in the inset in
Figure \ref{Fig_3}(a). Symmetry breaking often results in
weaker transmission probabilities in quantum transport,
and this has qualitative implications for the present
system: The symmetry breaking as energy levels are
shifted apart manifests itself in a  somewhat lower
transmission and current for parallel magnetization of
the contacts (especially at higher bias)  in Figure
\ref{Fig_3}(c) and (d)   than in Figure \ref{Fig_3}(a)
where the effect of the applied bias on the transmission
is neglected. However the net current for anti-parallel
magnetization is much less sensitive to bias-related
symmetry effects.\cite{antiparallel_trans}      Because
it selectively depresses the current for parallel
magnetization of the contacts,  the bias-induced symmetry
breaking results in a crossover with increasing bias from
positive to negative values of the JMR in Figure
\ref{Fig_3}(c) and (d), an effect not found in the less
realistic model of Figure
\ref{Fig_3}(a) where the effect of the bias on the
electron transmission probability is neglected. 

The bias voltage at which we predict negative JMR to
appear ($\sim$$2V$ in Figure
\ref{Fig_3}(c) and $\sim$$3.5V$ in Figure
\ref{Fig_3}(d)) depends on the details of the potential
profile across the junction where the two electrodes
touch:  In the linear voltage drop model the bias applied
across the junction $simultaneously$ shifts the energies
of similar atomic orbitals on the two tip atoms, bringing
them closer together or further apart, depending on their
relative spin orientations and whether the magnetizations
of the contacts are  parallel or antiparallel. Thus
orbitals of the tip atoms are brought closer to or
further from resonance with each other.   On the other
hand, since in the abrupt voltage drop model the applied
bias affects the energies  of all of the atomic orbitals
of the contacts $except$ those of the tip atoms, the
corresponding resonant effect of the applied bias in this
model is weaker: energies of orbitals within a cluster
are shifted with respect to {\em unshifted} energy levels
on the tip atoms (the energy levels of the rest of the
clusters are not directly coupled to any other atoms in
the opposing cluster).  Therefore, the onset of negative
JMR requires a higher applied bias in the abrupt voltage
drop model.   

\subsection{Spin-Dependent Transport for More General
Junction Geometries}

It has been established experimentally that the
conductance characteristics of atomic chains can be
altered by stretching them
\cite{Rubia1996}.  We now investigate the
corresponding dependence of the JMR in our magnetic
system.

\subsubsection{Generalizing the Tight-Binding Model}
\label{var geom}

For the more general junction geometries, our Hamiltonian
matrix elements between atomic orbitals within each of
the two Fe clusters will again be based, as discussed
above, on the Fe ferromagnetic tight-binding parameters
derived from {\em ab initio} bulk band structure
calculations\cite{Pap}. However  a different approach is
needed to obtain the non-diagonal Hamiltonian matrix
elements that describe tip-to-tip coupling since the
atoms involved no longer all fall on sites of a single
bcc Fe lattice. We estimate these matrix elements using
an appropriate modification of extended H{\"u}ckel
theory.  Extended H{\"u}ckel\cite{EH} is a semi-empirical
tight-binding model from quantum chemistry that provides
a simple description of the electronic structures of a
wide variety of molecules. It uses a non-orthogonal
atomic orbital basis. The diagonal Hamiltonian matrix
elements $H_{i,i}$ are identified with the experimental
ionization energies of the corresponding atomic orbitals
$i$ of isolated atoms. In the Wolfsberg-Helmholtz form of
the model\cite{Wolfsberg1952} the non-diagonal matrix
elements are assumed to be  
\begin{equation} H_{i,j} = \frac{K}{2} (H_{i,i} +
H_{j,j}) S_{i,j}
\label{HuckEQ}
\end{equation} where $S_{i,j}$ are the overlap matrix
elements connecting orbitals $i$ and $j$ that are
calculated by approximating the atomic orbitals $i$ and
$j$ with linear combinations of Slater-type orbitals.
$K$ is a phenomenological parameter, usually chosen to be
1.75, in order to match experimental data.  We estimate
the non-diagonal Hamiltonian matrix elements that
describe tip-to-tip coupling using equation
(\ref{HuckEQ})  and numerical values of the diagonal
Hamiltonian matrix elements based on those given in Ref.
\onlinecite{Pap}.  However, the values of the diagonal
matrix elements given in Ref. \onlinecite{Pap} are not
close to the ionization energies of the corresponding
atomic orbitals but are defined up to an arbitrary
additive constant. It is necessary to choose the value of
this constant with care in order to obtain realistic
results from equation (\ref{HuckEQ}). We choose its
value  so that when the two tips are positioned in the
geometry studied in Section
\ref{nearest neighbor} ,  our model (including  the
tip-to-tip coupling matrix elements obtained using
equation (\ref{HuckEQ})) reproduces the
$up \rightarrow up$ transmission near the Fermi energy
obtained using only the bulk parameters from Ref.
\onlinecite{Pap}. The value of the shift that we apply to
the bulk electronic parameters is $c =
-13.53eV$.\cite{non-diag shift}    The same tip-to-tip
coupling parameters $H_{m,n}$ are then used for all spin
configurations.\cite{more details}  This is consistent
with the fact that bulk $up$-$up$ coupling parameters are
very similar to $down$-$down$ parameters
\cite{Pap}. 

Thus in our calculations the overlaps $S_{i,j}$ between Slater-type 
orbitals provide the {\em distance and orientation-dependence}
of the tip-to-tip Hamiltonian matrix elements required for the 
more general junction geometries (as in extended H{\"u}ckel theory),
however, the Hamiltonian matrix elements are normalized so as to yield
transport results consistent with those obtained from a Hamiltonian
matrix derived from {\em ab initio} calculations.    

\subsubsection{Structural Considerations: Bulk and Relaxed Geometries}
\label{Structural Considerations}
  
The first geometry we consider using this hybrid model is
shown in the inset of Figure \ref{Fig_4}(a).  The two
(100) 5x5, 4x4, 3x3, 2x2, and single tip atom clusters
are now aligned with a common axis.  The two tip atoms
are again separated by a bulk nearest neighbor distance,
2.482{\AA}.  We refer to this as the ``linear geometry,"
and to the structure in the inset of Figure
\ref{Fig_3}(a) as the ``skewed geometry."  Our geometry
relaxations\cite{G03} on simple model systems involving
the pair of 2x2 planes and the two atoms forming the
dimer, indicate that the two tip atoms in this geometry
are close to their stable positions. Similar relaxations
that we carried out showed that when the clusters are pulled apart, some
stretching occurs between the tip atoms and   their respective 2x2 layers,
but most of the stretching occurs between the two tip atoms.  When the tips
are far apart representing vacuum tunneling, our geometry relaxations show
that, with respect to the rest of the cluster, the tip atom sits very near
its bulk position.  Therefore, we assume that as the junction is stretched,
all of the stretching occurs between the two tip atoms, and so the atoms of
each tip separately are located on their bulk Fe lattice
positions. Therefore, in what follows, we again make use the
bulk Fe ferromagnetic tight-binding parameters to describe the electronic
structure within the clusters for all matrix elements in $H$, and use our
modified H{\"u}ckel approach to describe the inter-cluster 
coupling.\cite{bulkg} 

\subsubsection{SDT in Fe Nanocontacts Bridged by Fe Atoms
in Contact in a Linear Geometry}

\begin{figure}
\includegraphics[width=0.99\columnwidth]{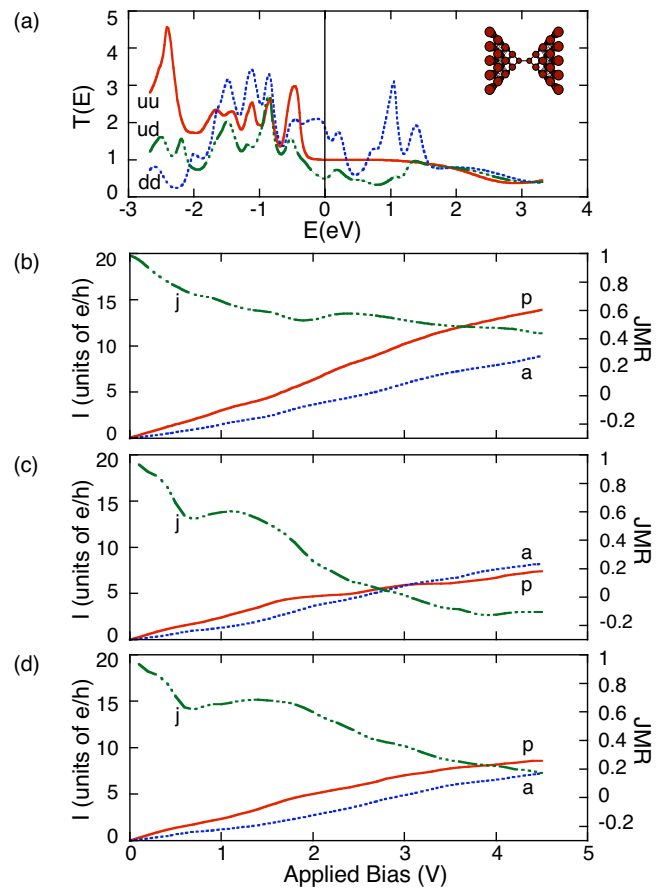}
\caption{(Color online) (a) The transmission probabilities as a
function of energy (eV) at zero bias for the spin $up
\rightarrow up$ (uu), $down
\rightarrow down$ (dd), and $up \rightarrow down$ (ud)
configurations for the Fe atoms at the nearest neighbor
separation, arranged in the linear geometry (inset).  The
Fermi energy is located at 0eV. (b) The current as a
function of voltage for parallel (p) and anti-parallel
(a) magnetizations in the zero bias approximation for the
transmission probabilities.  The JMR (j) is also shown.
(c) The currents and JMR as calculated with an applied
bias using the linear voltage drop model. (d) The
currents and JMR as calculated with an applied bias using
the abrupt voltage drop model.}
\label{Fig_4}
\end{figure}

Our results for spin transport in the linear geometry
with the tip atoms separated by a bulk nearest neighbor
distance are shown in Figure
\ref{Fig_4}.  The spin $up \rightarrow up$ transmission
shown in Figure
\ref{Fig_4}(a) shows a similar plateau above $E_{F}$ to
that in the skewed geometry (Figure \ref{Fig_3}(a)). 
This similarity is due to the dominance of the
non-directional
$s$-orbitals in both cases.  Below the Fermi energy, the
transmissions exhibit significant differences, due to the
importance of the very directional $d$-orbitals in that
energy range.  For the two different geometries, spin
$down \rightarrow down$ and $up
\rightarrow down$ transmissions are similar in a broad
sense, such as overall magnitude, but display many
differences on a finer scale due to the importance of $d$
orbitals in those transmissions.  Figure
\ref{Fig_4}(b) shows that the current through this
geometry, as calculated from the voltage-independent
transmissions, is smaller than that through the skewed
geometry using bulk parameters to describe the coupling.
However, with second nearest-neighbor couplings across
the junction turned off in the skewed geometry, the
current in the parallel magnetization configuration is in
very close agreement for the two geometries suggesting
that the lower current can be attributed to the lack of
second nearest neighbor coupling between the two contacts
in the linear geometry.  In the anti-parallel
magnetization configuration, the current is substantially
lower in the linear geometry than in the skewed geometry.
This reflects  the lower spin $up \rightarrow down$
transmission in the the linear geometry which results
partly from weaker
$d$ coupling: Many of the overlaps involving $d$ orbitals
are zero in the linear geometry.  The substantially lower
anti-parallel current results in a significantly stronger
JMR.

As shown in Figure \ref{Fig_4}(c) the current for
parallel magnetization and the magneto-resistance, as
calculated from the voltage-dependent transmission in the
linear voltage drop model, are again significantly lower
than for the voltage-independent transmission model for
the linear geometry (at higher bias) because the
Hamiltonian is no longer symmetric.  The current is
slightly weaker for the linear geometry than for the
skewed geometry and a larger JMR is predicted than for
the skewed geometry.  A negative JMR is again predicted,
but this time it manifests near a bias of
$3V$.  As was discussed for the skewed geometry, in the
linear voltage drop model the applied bias {\em
simultaneously} moves the atomic orbitals of the tip
atoms closer or further apart in energy. Thus the linear
voltage drop model provides a relatively strong mechanism
for those energy levels to move closer to or further from
resonance with each other.  However, in the present case
of the linear geometry, there is no second neighbor
coupling between a tip atom and atoms of the opposite
contact and so the resonant effect is weaker than in the
previous case of the skewed geometry.  The potential
profile with abrupt voltage drops is also less conducive
to negative JMR here (as in the skewed geometry) thus a
negative JMR does not occur within the voltage range in
Figure
\ref{Fig_4}(d).  It should also be noted that the
predictions regarding
$I-V$ characteristics and JMR in the two geometries are qualitatively similar for the linear voltage
drop model which, according to density functional calculations,\cite{Ke2004} may be a more accurate
approximation for all-metal systems than the abrupt voltage drop model. However, the predicted results
are qualitatively similar for {\em both} models of the  potential profile in both systems,
demonstrating the robustness of our method; it is reasonable to expect the results of a fully
self-consistent calculation of the potential profile to fall between these two model profiles. 
 Again, all of these results  are also reasonably robust to the addition of a 6x6  atom layer to each of
the clusters.

\subsubsection{Dependence of the Spin Transport on the
Separation Between Tip Atoms and Dangling Bond Formation}

\begin{figure}
\includegraphics[width=0.99\columnwidth]{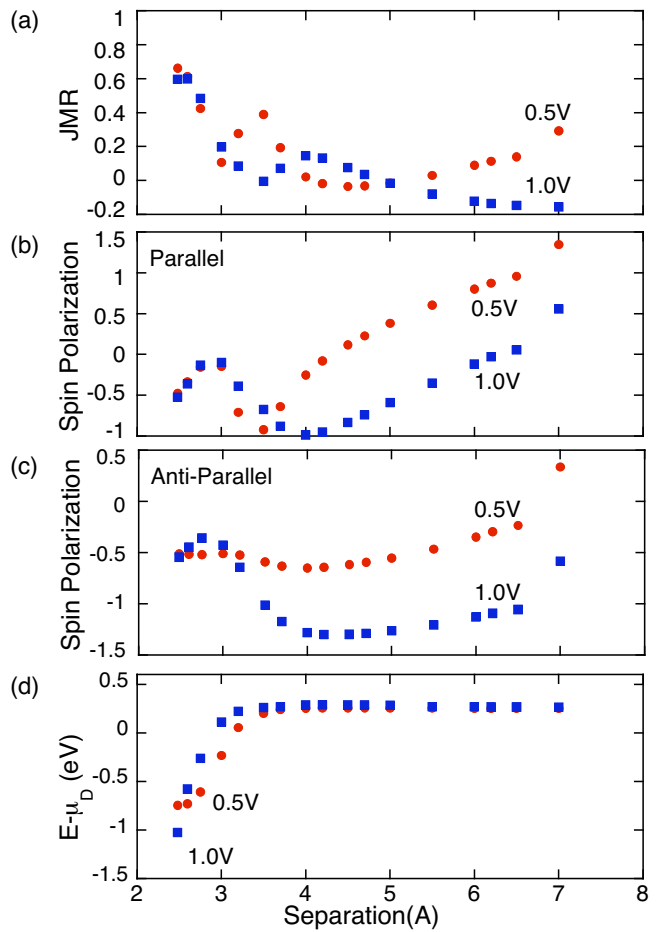}
\caption{(Color online) Spin dependent transport in the linear tip
geometry vs. separation of the tip atoms. (a) The JMR as
a function of tip separation calculated using the linear
voltage drop model and applied biases of 0.5V and 1.0V.
(b) The spin polarization of the current as a function of
tip separation for parallel magnetization of the
contacts. (c) The spin polarization of the current as a
function of tip separation for anti-parallel contact
magnetizations. (d) The energy eigenvalue associated with
the spin $down$ dangling bond state relative to the
chemical potential of the drain electrode (0eV) as a
function of tip separation for parallel magnetization of
the contacts.}
\label{Fig_5}
\end{figure}

Figure \ref{Fig_5}(a-c) shows the calculated dependence
of the magnetoresistance and spin polarization of the
current (as defined by Eqs. (\ref{JMR}) and (\ref{SP}),
respectively) on the separation between the two tip atoms
as the junction is stretched for two values (0.5 and
1.0V) of the bias voltage, for the linear voltage drop
model.  

As the separation between tip atoms is increased, the JMR
(Figure
\ref{Fig_5}(a)) initially decreases rapidly  then
increases to a local maximum at tip separations near 3.5
and 4{\AA} then resumes its decrease turning  weakly
negative and, for the $0.5V$ bias, increases again at
large separations. The spin polarization of the current
for parallel magnetization of the contacts (Figure
\ref{Fig_5}(b)) is negative at small separations, i.e.,
the spin $down$ (minority spin) current predominates, and
rises initially with increasing separation almost to zero
near 3{\AA}. It then passes through a minimum near 3.5
and 4{\AA} (where the JMR shows a maximum) before
resuming its rise and becoming positive at large
separations. The local maxima in the JMR and minima in
the spin polarization of the current near 3.5 and 4{\AA}
are due to a pronounced resonance in the dominant $down
\rightarrow down$ transmission that appears near the
Fermi energy for parallel magnetization of the contacts
at such separations. This transmission resonance persists
when a 6x6 Fe layer is added to each nanocontact, thus it
does not appear to be a dimensional resonance due to the
finite size of our Fe clusters.

By examining the contributions of the individual
electronic eigenstates of the Hamiltonian of the coupled
Fe clusters to the $down \rightarrow down$ transmission
(through the Green's function $G_{0}$ in the
Lippmann-Schwinger equation  (\ref{LSeq})) we have
identified the particular eigenstate that is responsible
for this transmission resonance. Our Mulliken analysis 
revealed that for tip separations corresponding to the
appearance of the $down \rightarrow down$ transmission
resonance, a significant portion of this eigenstate
resides on the tip atom of the drain contact
\cite{sourcestate} and has
$d(x^{2}-y^{2})$ orbital symmetry there; the
$d(x^{2}-y^{2})$ orbital has a lobe oriented along the
$x$-axis towards the tip atom on the other cluster. 
Since the
$d(x^{2}-y^{2})$ orbital is involved in tip-to-tip
bonding and the amplitude of the eigenstate of interest
on the tip atom increases as the tip separation is
increased, we attribute the appearance of this eigenstate
state to bond-breaking between the tip atoms and the
associated formation of a dangling bond.

The energy $E$ of the dangling bond eigenstate relative
to the electrochemical potential of the drain is plotted
in Figure
\ref{Fig_5}(d) as a function of tip separation for the
two values of applied bias considered in Fig.
\ref{Fig_5}(a-c).  The dangling bond state is the
dominant feature in the Mulliken spectra of the tip atom
within a broad window about the Fermi energy and this
criterion is used to identify the state for the different
values of the tip separation and applied bias.  At zero
temperature, only states within the limits of integration
corresponding to the applied bias contribute to the
predicted current (see Eq.
\ref{LandauerEQ}).  These limits are determined by the
electrochemical potentials of the source (0.5eV and 1.0eV
in Figure
\ref{Fig_5}(d) for the two values of applied bias) and
the drain (0eV).  As shown in the figure, as the tip
separation is increased, the energy of the dangling bond
state shifts into the window of integration, above the
electrochemical potential of the drain and therefore
begins to contribute to the current.  (Note that this
happens at separations close to those where the JMR in
Fig.\ref{Fig_5}(a) first begins to rise signaling the
onset of the transport resonance).   Simultaneously, the
Mulliken weight of the state located on the drain tip
increases as the separation is increased to a value of
about 10 percent, a sizeable portion of the total
probability distribution of the eigenstate given that a
total of 110 atoms make up the Fe clusters in our model. 
Therefore, as the separation is increased from its
smallest distance of 2.482{\AA}, we attribute the initial
decrease in JMR to the stretching of the bond between the
two tip atoms.  The following increase to the local
maximum in JMR is attributed to the formation of a
dangling bond as the tips are further pulled apart,
leading to the resonant feature in the $down
\rightarrow down$ transmission.  Once the dangling bond
has been formed, its energy and Mulliken weight are
roughly constant, and the JMR resumes its decrease with
increasing separation. 
 
As was discussed in Section \ref{nearest neighbor}, $s$,
$p$ and $d$ electrons all play a significant role in
transport in this system for small tip separations. 
However their contributions decay differently as the tip
separation is increased and the tunneling regime is
entered. Since the valence
$d$-electrons have a lower site energy than the other
electrons ($E_{d} < E_{s},E_{p}$),\cite{Pap} they decay
more rapidly resulting in a less significant contribution
from
$d$-electrons to transport as the tip separation is
increased.

For an applied bias of 0.5V and near a tip separation of
about 4.5{\AA}, the contribution to tunneling from
$d$-electrons is roughly equal to that of $s$-electrons. 
Here $T_{up
\rightarrow up}$, mostly due to
$s$-electrons, and
$T_{down \rightarrow down}$, mostly due to $d$-electrons,
are roughly equal and the spin polarization of the
current in the parallel magnetization case (Fig.
\ref{Fig_5} (b)) is roughly zero, defining a cross-over
regime.  This can be viewed as the cross-over from
ballistic transmission to a tunneling-like transport.  At
the cross-over, transmission from spin
$up$ (mostly $s$ and $p$) to spin $down$ (mostly $d$)
for  antiparallel magnetization of the contacts is
roughly the same as that for the parallel spin
configurations, and the resulting JMR (Fig. \ref{Fig_5}
(a), 0.5V) is very small, or even slightly negative. 
Since this small negative magneto-resistance appears at a
relatively small applied bias, it should be accessible
experimentally in these systems.    

At larger separations, beyond the cross-over regime, the
current is dominated by the transport of $s$ and $p$
electrons and $T_{up
\rightarrow up}$ makes the largest contribution. 
Therefore the spin polarization of the current in the
parallel magnetization case (Fig.
\ref{Fig_5} (b)) is positive, and growing.  Since
transmission involving
$down$ (mostly
$d$) electrons is rapidly decreasing,
$T_{down \rightarrow down}$ and $T_{up \rightarrow down}$
are small (due to band mismatch the anti-parallel
transmission in general is smaller than even the
decreasing $down \rightarrow down$ transmission) and, as
can be seen in Fig.\ref{Fig_5} (a) in the
$0.5V$ case, the JMR also begins to slowly grow.   At
higher biases (for example at 1.0V bias in
Fig.\ref{Fig_5}(a) and (b)) more energy levels are
sampled, increasing the importance of the $d$ states and
the cross-over doesn't occur until larger separations.
Therefore, this model predicts that when
$d$ states are important, the current will be dominated
by minority electrons, and when $d$ states are
negligible, current due to majority electrons will
dominate.  This is quite analogous to the predictions
made by Mathon on periodic systems involving Co and
tunneling gaps of varying widths, where the fast decrease
in
$d$-electron transport accounts for a rapid reversal in
sign of the spin-polarization \cite{Mathon2002}.

Since the application of bias breaks the symmetry of the
Hamiltonian,
$T_{up \rightarrow down} \ne T_{down
\rightarrow up}$ and a non-zero spin polarization of the
current  is predicted in the anti-parallel magnetization
case as well.  This effect is entirely due the voltage
drop across the junction and so is expected to be unique
to magnetic transport systems that can sustain a
significant potential drop across the junction, i.e,
atomic nanocontacts and tunnel junctions.  As shown in
Figure
\ref{Fig_5}(c), the spin polarization of the current for
antiparallel magnetization of the contacts (as defined by
Eq. \ref{SP}) is negative over a wide range of tip
separations due to the dominance of the $up
\rightarrow down$ transmission under applied bias as a
result of shifting energy levels.  Even at relatively low
bias, there is more source and drain energy level overlap
in this configuration than in the
$down \rightarrow up$ configuration where levels are
shifted apart by the applied bias.  At higher bias, this
is even more the case, resulting in a much more negative
spin polarization of the current at $1.0V$. 
Never-the-less, the spin polarization does increase for
large tip separations as the
$d$-state transport becomes less significant.  The
non-monotonic behavior of the spin polarization of the
current with tip separation for separations below 4{\AA}
in Fig.\ref{Fig_5}(c) is due to resonant transport
associated with dangling bond formation  similar to that
for the case of parallel magnetization of the contacts
that is discussed in detail above.  

\subsubsection{Vacuum Tunneling Between Fe Nanocontacts}

\begin{figure}
\includegraphics[width=0.99\columnwidth]{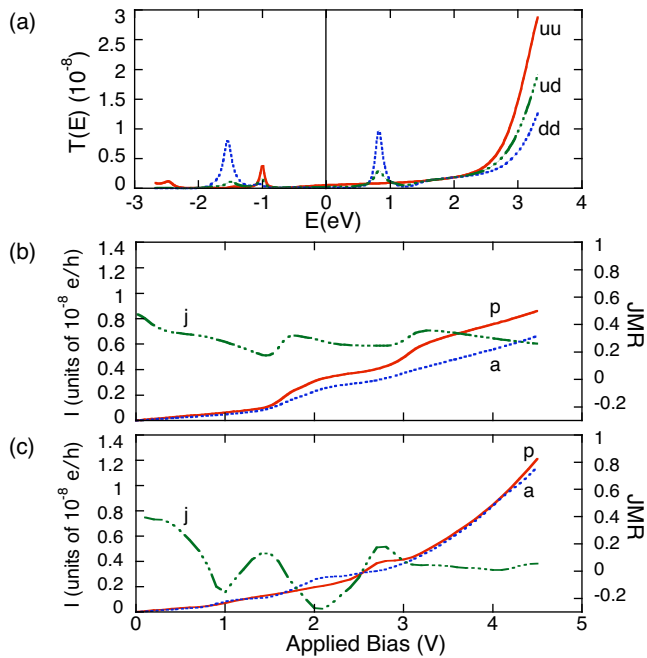}
\caption{(Color online) (a) The transmission probabilities as a
function of energy (eV) at zero bias for the spin $up
\rightarrow up$ (uu),
$down \rightarrow down$ (dd), and
$up \rightarrow down$ (ud) configurations for a 7{\AA}
tip separation representative of the vacuum tunneling
regime.  The Fermi energy is located at 0eV. (b) The
current as a function of voltage for parallel (p) and
anti-parallel (a) magnetizations as calculated by
integrating the zero bias transmission probabilities in
(a).  The JMR (j) is also shown. (c) The currents and JMR
as calculated with an applied bias using the linear
voltage drop model.}
\label{Fig_6}
\end{figure}

When the separation between tip atoms is very large
(vacuum tunneling), the $d$-electron states have decayed 
across the gap between the contacts and do not
significantly contribute to the transmission.  This
situation is easier to analyze as the couplings between
tip atoms involving $s$ and
$p_{x}$ (all the rest are negligible) behave simply. 
Figure
\ref{Fig_6}(a) shows the calculated transmission for the
linear geometry with a tip atom separation of 7{\AA} at
zero bias.  The currents calculated using the zero bias
approximation for $T$, shown in Figure
\ref{Fig_6}(b), display step-like features due to peaks
in the transmission.  The zero-bias approximation for $T$
also predicts a strong JMR effect, which shows broad
peaks where the transmission resonances are encountered.

Figure \ref{Fig_6}(c) shows the $I-V$ characteristics
calculated using the linear voltage drop model for
$T(E,V)$.  The application of bias and shifting of energy
levels strongly damps the local maxima in the zero bias
transmissions and the step-like behavior seen in Figure
\ref{Fig_6}(b) is now less pronounced.  The current
increases with bias, resembling the accepted behavior of
tunneling through vacuum.  At the application of about
$2V$, the $up
\rightarrow down$ transmission becomes the strongest
scattering channel, so the anti-parallel current becomes
stronger than the parallel current.  This results in a
relatively strong, negative JMR, although the
$down \rightarrow down$ transmission regains its
dominance and positive JMR returns at higher bias.  Thus
the shifting of different  transmission maxima again
results in non-monotonic behavior  of the JMR and
negative JMR for some values of the applied bias. 
However, there is no sustained JMR reversal such as is
predicted when $d$-electron states contribute to
transport.

\section{Conclusions}
\label{Conclusion}

We have presented a microscopic quantum theory of
spin-dependent transport across iron nanoscale junctions
bridged by chains of Fe atoms, based on the Landauer
approach to transport, semi-empirical tight-binding
Hamiltonians and Lippmann-Schwinger and Green's function
scattering techniques.  We first applied bulk
ferromagnetic tight-binding parameters (that were shown
to also provide a satisfactory description of surface
properties) to study ballistic transport between a pair
of Fe contacts connected by two Fe atoms in a nearest
neighbor geometry. We presented theoretical predictions
for the current-voltage characteristics of this system
for parallel and anti-parallel magnetizations of the
contacts and predicted that negative spin polarization of
the current should occur at low bias.  We also predicted
that the junction magnetoresistance of this system should
switch sign from positive to negative with increasing
bias. Next, we extended our model so as to allow us to
study spin-dependent transport for more general tip
geometries and presented predictions of the junction
magnetoresistance and spin polarization of the electric
current through Fe nanoscale junctions as a function of
separation between the tip atoms of the nanocontacts from
nearest neighbor distances to the vacuum tunneling
regime. Characteristic trends emerging in those transport
predictions as the separation is varied were associated
with decay rates of different orbitals and with the
breaking of bonds and associated dangling bond formation
resulting in spin-dependent transmission resonances.  We
also presented a systematic physical interpretation of
our  predictions that in many cases are qualitatively
different from those that rely on density of states
considerations alone, and also different from the
behavior of SDT in thin film junctions where lattice
periodicity in directions orthogonal to that of the
current flow plays a key role.  Many of our predictions
apply in the regime of low bias that should be accessible
with present day experimental techniques.     

\section*{Acknowledgments}

We thank B. Heinrich, R. Hill and B. L. Johnson for
helpful discussions. This research was supported by NSERC
and the Canadian Institute for Advanced Research.


\begin{thebibliography}{}
\bibitem{zeroT}  The calculations presented here were
carried out for zero temperature.
\bibitem{non-diag shift}It should be noted that because
the orbital basis is non-orthogonal, if the diagonal
matrix elements of the bulk tight-binding Hamiltonian of
Ref. \onlinecite{Pap}  are shifted by $c$ the
non-diagonal elements must be shifted by $cS_{i,j}$ for
consistency.\cite{shift}
\bibitem{more details}The transmission away from the
Fermi energy for the spin $up$ configuration with this
parameterization is in good qualitative agreement with
that found utilizing the bulk parameters for the tip-tip
coupling.  The spin $down$ and anti-parallel
configurations also yield transmission in qualitative
agreement near and away from the Fermi energy.  This
scheme succeeds in fitting the total transmission, and
independently, the portion of the transmission due to $d$
orbital coupling, and the combined transmission due to
$s$ and $p$ orbitals, however the resulting transmission
has less $p$ character than for the bulk
parameterization.
\bibitem{bulkg} Note that our use of bulk geometries and electronic 
parameters for the atoms of the clusters other than the tip atoms is justified by the
fact that the Fe clusters in our model represent the ends of macroscopic
bulk Fe leads.
\bibitem{sourcestate} A similar eigenstate with enhanced
weight on the tip atom of the source contact is also
found but under most conditions its energy is outside of
the range between the source and drain electrochemical
potentials so that it does not contribute significantly
to the current.
\end{thebibliography}
\end{document}